\documentclass[conference]{IEEEtran}
\IEEEoverridecommandlockouts

\usepackage{amsmath,amssymb,amsfonts}
\usepackage{comment}
\usepackage{ulem}
\usepackage{cite}
\usepackage{framed}
\usepackage{color}
\usepackage{listings}
\usepackage{latexsym}
\usepackage{graphicx}
\usepackage{url}

\usepackage{algorithm}
\usepackage{algpseudocode}

\begin{document}

\title{Locating Buggy Segments in Quantum Program Debugging
}

\author{\IEEEauthorblockN{Naoto Sato and Ryota Katsube}
\IEEEauthorblockA{\textit{Research \& Development Group, } 
\textit{Hitachi, Ltd.}\\
naoto.sato.je@hitachi.com}

}

\maketitle

\begin{abstract}
When a bug is detected by testing a quantum program on a quantum computer, we want to determine its location to fix it. To locate the bug, the quantum program is divided into several segments, and each segment is tested. However, to prepare a quantum state that is input to a segment, it is necessary to execute all the segments ahead of that segment in a quantum computer. This means that the cost of testing each segment depends on its location. We can also locate a buggy segment only if it is confirmed that there are no bugs in all segments ahead of that buggy segment. Since a quantum program is tested statistically on the basis of measurement results, there is a tradeoff between testing accuracy and cost. Although these characteristics are unique to quantum programs and complicate locating bugs, they have not been investigated. We suggest for the first time that these characteristics should be considered to efficiently locate bugs. We are also the first to propose a bug-locating method that takes these characteristics into account. The results from experiments indicate that the bug-locating cost, represented as the number of executed quantum gates, can be reduced with the proposed method compared with naive methods.
\end{abstract}

\section{Introduction}
The field of quantum software engineering has developed rapidly \cite{piattini2020talavera}\cite{piattini2021quantum}\cite{survey5}\cite{moguel2020roadmap}\cite{zhao2020quantum}\cite{DESTEFANO2022111326}.
Research on testing, verification, and debugging of quantum programs began with a typology of bugs \cite{huang2018qdb} \cite{bugpattern}. On the basis of these studies, an application of classical methods to quantum programs was proposed \cite{classicalTest}. 
When a bug is detected by testing a quantum program on a quantum computer, we divide the program into several segments and test each segment to determine the detailed bug location. To find the bug in the testing of each segment, it is necessary to prepare the actual quantum states that would be input to each segment when the entire quantum program is executed. In a quantum computer, to prepare the actual quantum state for a segment, all segments ahead of that segment should be executed on the initial state that the quantum computer physically forms \cite{Long2022lwc}.
This leads to the cost of testing segments to depend on the locations of the segments. Even if a bug is detected in the testing of a segment, it does not necessarily mean that the bug is in that segment. This is because the bug may be in other segments ahead of that segment. 
Therefore, to locate a buggy segment on a quantum computer, we have to confirm that there is no bug in any segment ahead of it. As another perspective, the testing of each segment is conducted on the basis of measurements. Since a sufficient number of measurements is necessary for testing with sufficient accuracy, there is a tradeoff between testing accuracy and its cost.

These characteristics unique to quantum programs complicate locating bugs; however, they have not been mentioned in previous studies. The first contribution of this paper is clarifying for the first time the characteristics that should be considered to efficiently locate a buggy segment in quantum programs on a quantum computer. The second contribution is that we propose a bug-locating method for quantum programs. We implement the proposed method and conducted experiments to demonstrate its efficiency.

\section{Background}\label{background}
\subsection{Quantum Program Testing}\label{test}
\begin{figure}[b]
\centering
\scalebox{0.28}{\includegraphics[bb=0 0 735 196]{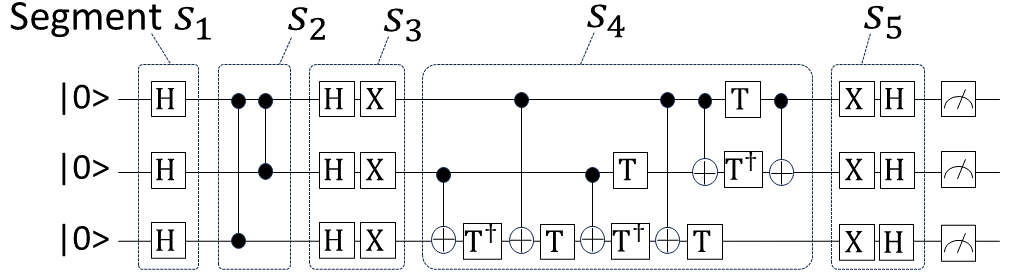}}
\caption{Quantum program divided into segments}
\label{unit}
\end{figure}

A qubit, which is a variable in a quantum program, is in a superposition of the basis state |0> and |1>. When a qubit is measured, 0 or 1 is observed probabilistically, depending on its superposition state. The state of a qubit can be expressed as $|\psi> = a_0 |0> + a_1 |1>$ ($|a_0|^2 + |a_1|^2=1$), where $a_0$ and $ a_1$ are complex numbers and called amplitudes. The absolute squares of the amplitudes $|a_0|^2$ and $|a_1|^2$ represent the probabilities of obtaining 0 and 1 by measurement. An arbitrary quantum state consisting of $n$ qubits is generally represented by $2^n$ basis states.
Figure \ref{unit} is an example of a quantum program implementing Grover's algorithm. The quantum program is represented by a model called quantum circuit. Each horizontal line corresponds to a qubit, and the operations on them, i.e., the quantum gates, are arranged from left to right.
We assume a quantum program so large that it cannot be simulated in practical time on a classical computer. Since the motivation for using a quantum computer is to solve complex problems that cannot be solved with a classical computer, this assumption is natural.

Long et al. classified testing methods for quantum programs into statistic-based detection (SBD) and quantum runtime assertion (QRA) \cite{Long2022lwc}. With SBD methods, the output quantum state is measured many times. Quantum information derived from the measurement results is then statistically compared with the test oracle. If they are different, the quantum program is statistically determined to have a bug. A simple SBD method is comparing the absolute square of the amplitude that can be directly derived from the measurement results \cite{huang2019statistical}. 
For a more rigorous test, the density matrix may be useful, which is calculated by quantum state tomography, maximum likelihood estimation, or Bayesian estimation \cite{nielsen2010quantum}\cite{TANAKA20122471}\cite{lukens2020practical}. 
Since QRA methods were proposed for runtime testing, they do not destroy the tested quantum state and can be conducted using the result of a single measurement \cite{li2020projection}\cite{liu2020quantum}\cite{runtime2021}. 
However, QRA methods require the embedding of a complex extra quantum circuit in the tested circuit. The implementation of the extra circuit is not easy, and if there is a bug in the extra circuit, the test cannot be conducted correctly \cite{Long2022lwc}. Therefore, if we do not want to run the tests at runtime, both SBD and QRA methods are useful.
Our proposed method for locating a buggy segment uses an SBD method internally, but it does not depend on the details of the testing method; thus, we used the testing method of comparing the absolute square of the amplitude in our experiments discussed in Section \ref{exp}.

\subsection{Related Work} \label{relwork}
Various SBD methods have been proposed by applying classical testing methods to quantum programs, such as fuzzing \cite{wang2021poster}, mutation testing \cite{mendiluze2021muskit} \cite{fortunato2022qmutpy}, search-based testing \cite{wang2022qusbt} \cite{wang2022mutation}, combinatorial testing \cite{wang2021application}, coverage testing \cite{assessing} \cite{wang2021quito}, metamorphic testing \cite{abreu2022metamorphic}\cite{MorphQ}, property-based testing \cite{honarvar2020property}, and equivalence class testing \cite{Long2022lwc}. Muqeet et al. proposed leveraging machine learning to improve the accuracy of testing under the noise effect \cite{muqeet2023noise}. We used a statistical assertion method that is based on the chi-square test in our experiments, which was proposed by Huang et al. \cite{huang2019statistical}. Li et al. presented runtime assertion as a QRA method using a projective measurement that stabilizes the tested quantum state \cite{li2020projection}.
Liu et al. introduced another runtime assertion by adding extra (ancilla) qubits to collect the information of the tested quantum state \cite{liu2020quantum}\cite{runtime2021}. 

Focusing on debugging quantum programs, Miranskyy et al. discussed the applicability of the classical debugging strategies to quantum programs \cite{miranskyy2020your}\cite{miranskyy2021testing}. Liu et al. suggested using the information obtained from assertion tests for debugging \cite{liu2020quantum}. 
Zhao et al. introduced bug patterns of quantum programs and a static analysis tool based on the patterns \cite{zhao2023qse}. Li et al. proposed a debugging scheme to locate bugs by injecting assertions into a quantum program \cite{li2020projection}. They also suggested that to show that a segment has a bug, it is necessary to confirm that all segments ahead of it do not have a bug. However, it was not recognized as a factor that complicates bug locating, and their search strategy was based on naive linear search. This paper clarifies four characteristics that should be considered to locate bugs and presents our efficient bug-locating method that takes these characteristics into account.

\section{Characteristics of Quantum Program Testing for Bug Locating}\label{problem}
When a bug is detected in a quantum program, we divide the program into several segments and test each segment to determine the bug location. In testing each segment, we assume that the developer can expect the correct output quantum state of each segment as a test oracle. For example, the output state of segment $s1$ in Figure \ref{unit} is expected to be a uniform superposition state. Since the phases of some bases should only be inverted in $s2$, the same absolute square of the amplitude in the Z-basis as in $s1$ is expected as the oracle.
When the quantum circuit is large and complex, it is not always possible to give oracles for arbitrarily divided segments. In such cases, segments can be defined to correspond to logical units such as subroutines.

To find the bug in the test of each segment, the actual quantum states that would be input to each segment when the entire quantum program is executed should be used as test input. However, unlike a classical computer, a quantum computer cannot easily prepare the quantum state as desired. If we want to prepare the actual input state of a segment, all segments ahead of the segment should be executed. This means that when testing a backward segment, more quantum gates are executed than when testing a forward segment. That is, the first characteristic \textbf{[C1]} is that the cost of testing a segment depends on its location. It is also implied that if we test a segment and a bug is detected, it is possible that the bug is not in the segment but in another segment ahead of it. Therefore, the second characteristic \textbf{[C2]} is that we can locate a buggy segment only if it is confirmed that there is no bug in all segments ahead of the buggy segment.
This suggests another characteristic. If a bug is not detected in the test of a segment, it can be assumed that the segment ahead of it is also bug-free (more precisely, no bug affecting the output of the segment). Conversely, if a bug is detected, the same bug should also be detected in the tests of the segments behind it. Accordingly, the third characteristic \textbf{[C3]} is that the test results of the segments are not independent.

As described in Section \ref{test}, the test of each segment is statistically conducted on the basis of the measurement results. A sufficient number of measurements is necessary for testing with sufficient accuracy. This indicates the fourth characteristic \textbf{[C4]} that there is a tradeoff between testing accuracy and its cost. In the chi-square test that we used in our experiments discussed in Section \ref{exp}, accuracy is indicated by significance level and power. 
The power of the test described as $1 - \beta $ is the probability of correctly detecting the presence of a bug, where $ \beta $ is the Type I\hspace{-1.2pt}I error rate. The power depends on the reliability of a sample, which is represented by the standard error of the sample mean $ \sqrt{ \sigma^2 / M} $, where $ \sigma^2$ is the unbiased estimate of the population variance and $M$ is the sample size \cite{helie2007understanding}. Since the power depends not only on the sample size but also on the significance level, which corresponds to the Type I error rate, the significance level also affects the required sample size. This means that test accuracy depends on the number of samples, that is, measurements.
Although these four characteristics affect the efficiency of locating bugs, they have not been mentioned in previous studies.

\section{Proposed Method}\label{proposed}
The proposed method consists of {\it cost-based binary search}, {\it early determination}, {\it finalization}, and {\it looking back}.

\subsection{Cost-based Binary Search}\label{sec-tree}
Binary search is an efficient array search algorithm \cite{williams1976modification}, in which the location of the target value is recursively narrowed down by comparing the middle element of the array and target value. By selecting the center element of the array as the middle element, binary search ensures that the search costs of the left and right subarrays are as similar as possible.
We believe that binary search is also effective in locating buggy segments in quantum programs. However, as [C1] states, the cost of the test depends on the position of the segment. 
Let $S_l$ be a sequence of segments of length $l$. If a segment $s_x$ ($1 \leq x \leq l-1$) is the middle element, the segment sequences from $s_1$ to $s_x$ and from $s_{x+1}$ to $s_l$ are called the left sequence and right sequence in terms of $s_x$, respectively. If the central segment is selected as the middle element, the search costs of the left and right sequences are not similar. Therefore, we use {\it cost-based binary search} in which the middle element is selected on the basis of the testing cost.

\begin{figure}[hbt]
    \centering
    \scalebox{0.45}{\includegraphics[bb=0 0 471 258]{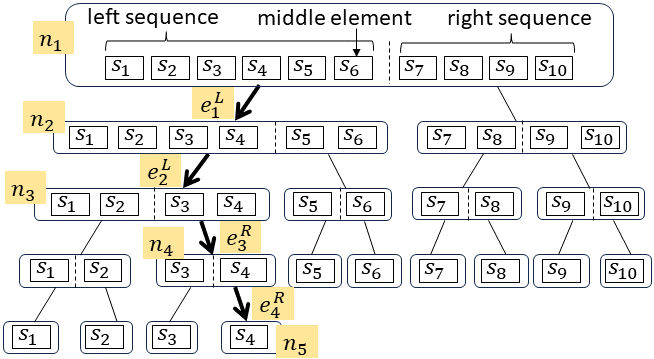}}
    \caption{Example of cost-based binary search tree}
    \label{tree}
\end{figure}

Since computational resources are expended for each gate execution, we define the testing cost $c_x$ as the number of quantum gates to be executed in the test of segment $s_x$. That is, $c_x = \sum_{i=1}^{x} g_i$, where $g_i$ denotes the number of quantum gates in $s_i$.
The middle element $s_x$ is selected so that the highest total testing costs for searching the left and right sequences are as similar as possible. 
Therefore, the index $x$ of middle element $s_x$ is given as
\begin{eqnarray*}
\label{smid}
\mathop{\arg \min}_{1 \leq x \leq l-1} \left| \sum_{i=1}^{x-1} c_i - \sum_{i=x+1}^{l-1} c_i \right|.
\end{eqnarray*}
An example of a cost-based binary search tree is shown in Figure \ref{tree}.
A node in the tree is associated with a segment sequence to be searched, which we call a target sequence, and the output state of the middle element is tested (represented with the dashed line). Note that when testing a target sequence, all segments ahead of the middle element should be executed, not the target sequence. We call this an executed sequence. In the example in Figure \ref{tree}, at node $n_4$, the target sequence consists of $s_3$ and $s_4$, and the executed sequence is from $s_1$ to $s_3$. If a bug is detected from the test at node $n_i$, move to the left node with the edge $e_i^L$. Otherwise, transit to the right node with the edge $e_i^R$.

\subsection{Search Strategies}\label{strategy}
\subsubsection{{\bf Early Determination}}\label{early}
In accordance with the binary search tree, we search for a buggy segment. On the basis of [C4], we introduce a strategy to reduce the search cost, which is called {\it early determination}. This is based on the assumption that statistically sufficient accuracy is over-performance for proceeding the search, and it may be more efficient to reduce the number of measurements by taking the risk of return in the search. As another motivation of {\it early determination}, we focus on the "reinforcement" relation between determinations, which is based on [C3].
A search path from node $n_1$ to $n_k$ is denoted as a sequence of edges $[e_1^{d_1}, ... , e_i^{d_i}, e_{i+1}^{d_{i+1}}, ... , e_{k-1}^{d_{k-1}}]$. Assume that there is no bug in the executed sequence $S_i$ of $n_i$, which corresponds to the null hypothesis of the test at $n_i$. The Type I error rate of the test at each node is denoted as $ \alpha $. If we determine at $n_i$ that there is a bug in $S_i$, that is $d_i=L$, the probability of making this determination under the null hypothesis is $ \alpha $.
In accordance with the structure of the search tree, the executed sequence $S_{i+1}$ of $n_{i+1}$ is included in the executed sequence $S_i$ of $n_i$. This means that there is also no bug in $S_{i+1}$ under the null hypothesis.
Therefore, if we determine $d_{i+1}=L$ at $n_{i+1}$, the probability is also $ \alpha $ under the null hypothesis of $n_i$. Finally, the probability of determining $d_i=d_{i+1}=L$ is $ \alpha ^2 $. This means that by proceeding from $n_i$ to $n_{i+1}$, the null hypothesis of $n_i$ can be rejected with more certainty at $n_{i+1}$ than at $n_i$, and rejecting the null hypothesis corresponds to the determination of $d_i=L$. We call this relation $e_{i+1}^L$ reinforces $e_i^L$. In the example in Figure \ref{tree}, $e_2^L$ reinforces $e_1^L$. The same is applied for $d_i = d_{i+1} = R$. The fact that past determinations may be confirmed later motivates {\it early determination}. However, an incorrect determination is not reinforced later; thus, we introduce strategies to modify the incorrect determination in Sections \ref{finalization} and \ref{suspicious}.

\subsubsection{{\bf Finalization}}\label{finalization}
{\it Early determination} is based on the assumption that sufficient accuracy is not necessary when advancing the search. When finally locating a buggy segment, however, the test should be conducted with sufficient accuracy.
Therefore, on the basis of [C2], the proposed method executes {\it finalization} when the binary search reaches a leaf node and locates the buggy segment $s_x$. {\it Finalization} consists of the tests of $s_{x-1}$ and $s_x$ with sufficient accuracy. It should be confirmed that the segments from $s_1$ to $s_{x-1}$ do not contain a bug from the test of $s_{x-1}$, and that the segments from $s_1$ to $s_x$ does contain a bug from the test of $s_x$, with sufficient accuracy. If {\it finalization} reveals an incorrect determination at a node, the search will return to the node.

\subsubsection{{\bf Looking Back}}\label{suspicious}
In addition to {\it finalization}, we introduce {\it looking back} to modify incorrect determinations. First, we show that we only need to focus on the last $L$ edge in a search path if we have incorrectly determined that there is a bug. Assume that the binary search is executed from $n_1$ to $n_k$ with the path $p=[e_1^{d_1}, ..., e_h^L, ..., e_i^L, e_{i+1}^R, ..., e_{k-1}^R]$ in which $e_h^L$ is an arbitrary $L$ edge from $n_1$ to $n_i$ ($1 \leq h \leq i-1$) and $e_i^L$ is the last edge of $L$. 
Since the executed sequence $S_i$ of $n_i$ is included in the executed sequence $S_h$ of $n_h$, if there is no bug in $S_h$, there is also no bug in $S_i$. That is, if $e_h^L$ is incorrect, $e_i^L$ is also incorrect. 
Therefore, when we want to confirm that there is an incorrect $L$ edge in the path from $n_1$ to $n_i$, we only need to check whether the last $L$ edge, $e_i^L$, is incorrect.
Next, we focus on the successor $e_{i+1}^R, ..., e_{k-1}^R$. If this is long, it suggests that $e_i^L$ is likely to be incorrect. If $e_i^L$ is incorrect, that is, $S_i$ does not include a bug, the executed sequences $S_{i+1}, ..., S_{k-1}$ also do not include a bug. In that case, $R$ edges appear in succession if the determinations are correctly executed.
Therefore, if $R$ edges appear more than a certain number of times in succession, we should suspect that $e_i^L$ is incorrect. {\it Looking back} confirms the determination of the last $L$ edge by the test with sufficient accuracy. If the determination turns out to be incorrect, the search returns to the node. 
Similarly, if $L$ edges appear in succession, the last $R$ edge is looked back to.

\section{Experiments}\label{exp}
\begin{table*}
\centering
\scalebox{0.6}{\includegraphics[bb=0 0 1786 190]{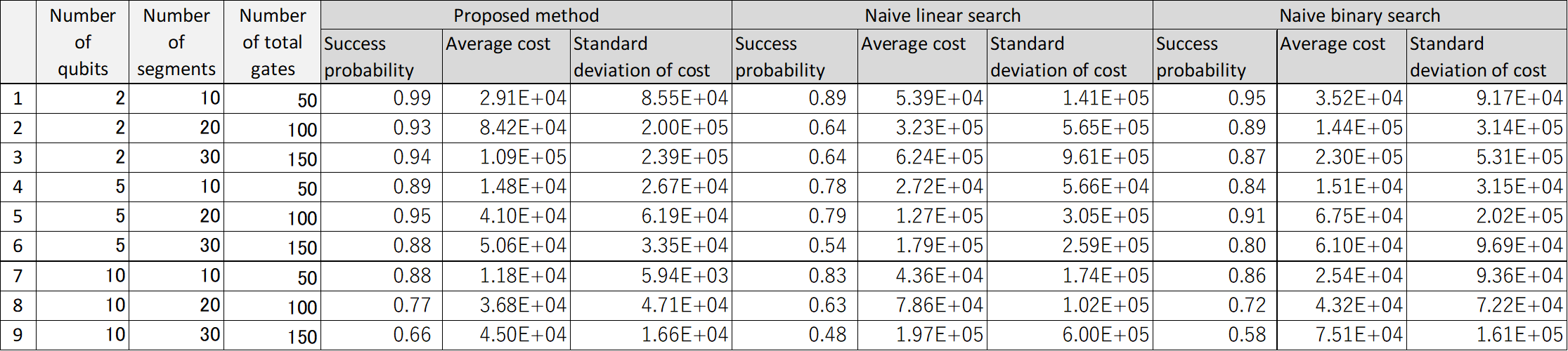}} 
\caption{Experimental results}
\label{table01}
\end{table*}
We implemented our proposed method and applied it to arbitrarily generated quantum programs. 
In each program, a bug was injected into a segment by arbitrarily replacing a quantum gate with another. In testing each segment, as in a previous study \cite{huang2019statistical}, we used the chi-square test, which statistically compares the absolute square of the amplitude with its oracle from the results of Z-basis measurements. Therefore, programs with bugs that cannot be detected by Z-basis measurement were excluded. We also implemented naive linear search and naive (non-cost-based) binary search in which the central segment is selected as the middle element, as comparison methods. With these naive methods, the chi-square test is also used for testing segments with sufficient accuracy, but the strategies of our method described in Section \ref{proposed} are not applied. In the chi-square test, the p-value and power of the test are referred to as accuracy indicators. When determining that there is a bug with sufficient accuracy, the thresholds of the p-value and power are 0.05 and 0.8, respectively. When applying {\it early determination}, only the p-value is referred to and its threshold is relaxed to 0.1. Similarly, when determining the absence of a bug, the threshold of the p-value with sufficient accuracy is set to 0.8, and it is relaxed to 0.6 in {\it early determination}. The power is not referred to when determining the absence of a bug because it is similar to the significance level when there is no bug. {\it Looking back} is executed when the same kind of edge ($R$ or $L$ edge) appears three times in succession.
At each node, measurements are repeated until these indicators exceed the thresholds, but an upper limit is defined for the number of measurements. If the upper limit is reached, the search fails. The experiment was conducted through simulation on a classical computer using Qiskit\textregistered \cite{cross2018ibm}.
The experimental results are listed in Table \ref{table01}. By generating 100 quantum programs for each row, we evaluated the probability and average cost (the average number of quantum gates executed) of successfully locating a buggy segment. We also calculated the standard deviation of the cost. 

Table \ref{table01} shows that the costs and standard deviations of our method are lower than those of the naive methods. The results of the standard deviations indicate that the search costs are more equalized by {\it cost-based binary search}. The results also indicate that the success probabilities of the proposed method are higher than those of the naive methods. If the difference in the output state caused by the bug is small, it is difficult to determine the presence of the bug by the test with sufficient accuracy. Therefore, it is likely that the number of measurements reaches the upper limit and the search fails there. Since the proposed method uses {\it early determination}, the search is more likely to proceed before the number of measurements reaches the upper limit than with the naive method. The experimental results indicate that {\it early determination} also contributes to the improvement of the success probabilities.

\section{Discussion}\label{discuss}

The basic idea of {\it early determination} is to take the risk of return instead of reducing the number of measurements. This section discusses the probability of return on the basis of Bayes' theorem. 
Assume that the quantum program is divided into $l$ segments and the search is executed from $n_1$ to $n_k$ with the path $p=[e_1^{d_1}, ..., e_i^{d_i}, ..., e_{k-1}^{d_{k-1}}]$. At node $n_i$, the segment $s_x$ ($1 \leq x \leq l-1$) is tested, and the executed sequence is $S_i$.
When {\it early determination} is applied at each node, the Type I and Type I\hspace{-1.2pt}I error rates of a statistical test are denoted as $ \alpha $ and $ \beta $, respectively.

The case $d_i=L$ is described as follows, but the same is applied for $d_i=R$. First, we consider the prior probability $P(B)$ that $S_i$ does not have a bug. For sake of simplicity, we assume the program contains only one bug and each segment has an equal chance of containing it. The $P(B)$ is expressed as $(l-x)/l$, where $x$ corresponds to the length of $S_i$. 
Next, let $w$ ($w \geq 1$) be the number of $L$ edges from $e_i^L$ to $e_{k-1}^{d_{k-1}}$. The conditional probability $P(A|B)$ that the search follows path $p$ from $n_i$ to $n_k$ when $S_i$ contains no bug is expressed as $(\alpha)^w (1 - \alpha)^{k-i-w}$.
The marginal probability $P(A)$ that the search reaches $n_k$ from $n_i$ along $p$ is then calculated as a sum of the probabilities for each bug location. For sake of simplicity, instead of the actual $P(A)$, we use the largest term $(1 - \beta)^w (1 - \alpha)^{k-i-w}$ in $P(A)$, which is the probability that the buggy segment is correctly narrowed down at $n_{k}$. Finally, the posterior probability $P(B|A)$ that the search returns from $n_k$ to $n_i$ is expressed as
\begin{eqnarray*}
P(B|A) \simeq \frac{(\alpha)^w  (1 - \alpha)^{k-i-w}  ((l-x)/l)}{(1 - \beta)^w (1 - \alpha)^{k-i-w}}.
\end{eqnarray*}
Since $ \alpha $ is much smaller than $1$, focusing on $(\alpha)^w$, we see that $P(B|A)$ decreases exponentially as $w$ increases. This indicates that the presence of $L$ edges before $e_i^L$ decreases the probability of returning to $n_i$. In Section \ref{early}, we interpreted this as the reinforcement relation of edges, which is the basis for {\it early determination}. If $w=1$, that is, only $R$ edges appear after $e_i^L$, the probability to return to $n_i$ does not decrease. In this case, the correctness of $e_i^L$ will be checked by additional tests according to {\it looking back}.

\section{Conclusion and Future Plans}

We presented for the first time four characteristics that should be considered to locate buggy segments of quantum programs on a quantum computer. We also proposed an efficient bug-locating method consisting of {\it cost-based binary search}, {\it early determination}, {\it finalization}, and {\it looking back}. We also experimentally demonstrated the efficiency of the proposed method. 

Future plans includes conducting experiments with real and large quantum programs on an actual quantum computer. It is also necessary to apply each strategy separately to evaluate its efficiency. We will also demonstrate the usefulness of the proposed method in the entire debugging process, e.g., whether the proposed method can locate multiple bugs. The proposed method should locate the most forward buggy segment. By applying the proposed method again after fixing the bug, another segment including a bug will be located.
Improvements to the proposed method is also included for future plans. More appropriate thresholds of the accuracy indicators used in {\it early determination} can be theoretically calculated on the basis of the risk of return described in Section \ref{discuss}. 
The testing method we used for this study is measuring quantum states in the Z-basis and comparing the absolute square of the amplitude with its oracle by the chi-square test. 
Bugs that do not appear as a difference in the amplitude of the Z-basis, such as a difference in phase, cannot be detected. Therefore, we should consider leveraging measurements in different bases. The efficiency of the proposed method when used with other test methods described in Section \ref{relwork} should also be evaluated.

\bibliographystyle{ACM-Reference-Format}
\bibliography{debugnavi}

\end{document}